\documentclass{aa}  

\usepackage{graphicx}
\usepackage{txfonts}
\usepackage{lipsum}
\usepackage{subcaption}         
\usepackage{lscape}             
\usepackage{placeins}           

\usepackage{CJKutf8}
\usepackage{xcolor}

\newcommand{\cjk}[1]{{\begin{CJK*}{UTF8}{bsmi}#1\end{CJK*}}}

\begin{document}

   \title{Why Do Stars Turn Red?}

   \subtitle{II. Steady-State Envelope Solutions}

%

   \author{Po-Sheng Ou (\cjk{歐柏昇})\inst{1,2}\fnmsep\thanks{Corresponding author: psou@asiaa.sinica.edu.tw}
        \and Ke-Jung Chen (\cjk{陳科榮})\inst{1,3}
        }

   \institute{Institute of Astronomy and Astrophysics, Academia Sinica, No.1, Sec. 4, Roosevelt Rd., Taipei 106319, Taiwan, R.O.C.
   \and Department of Physics, National Taiwan University, No.1, Sec. 4, Roosevelt Rd.,  Taipei 106319, Taiwan, R.O.C.
   \and Heidelberg Institute for Theoretical Studies, Schloss-Wolfsbrunnenweg 35, Heidelberg 69118,
Germany}


 
  \abstract
  {}
   {The physical origin of red giants (RGs) and red supergiants (RSGs) remains a fundamental question in stellar astrophysics. In Paper II of this series, we investigate the physical mechanisms governing envelope expansion toward the RG/RSG phase by systematically exploring the physically realizable configurations of stellar envelopes.} 
   {We construct steady-state stellar envelope models by solving the time-independent stellar structure equations while neglecting the core. The inner boundary is defined by a fixed pressure condition motivated by MESA stellar evolution models presented in Paper I.}
   {Our models show three key features of envelope expansion toward the RG/RSG phase.
(1) The refined mirror principle identified in Paper I is recovered: the post-main-sequence stellar radius varies inversely with the radius of the envelope’s inner boundary, arising purely from hydrostatic equilibrium.
(2) We identify an upper limit to envelope expansion corresponding to an effective temperature of $\sim 4{,}000\,{\rm K}$, characteristic of RG/RSG stars and consistent with the Hayashi limit. This temperature limit is regulated by H$^{-}$ opacity, whose sharp decline at low temperatures flattens the surface temperature gradient and drives a structural transition.
(3) The yellow regime of intermediate radius corresponds to an instability zone, in which small displacements of the hydrogen-burning shell produce large variations in stellar radius, naturally accounting for the bifurcation of giants and supergiants into blue and red branches instead of remaining in the yellow regime.}
{}


   \maketitle
   \nolinenumbers

\section{Introduction} \label{sec:intro}

The physical origin of red giants (RGs) and red supergiants (RSGs) remains a fundamental problem in stellar astrophysics. In Paper~I of this series \citep{Ou2024}, we investigated the mechanisms driving post-main-sequence expansion using stellar evolution models computed with the Modules for Experiments in Stellar Astrophysics code \citep[MESA;][]{Paxton2011,Paxton2013,Paxton2015,Paxton2018,Paxton2019,Jermyn2023}. We demonstrated that explanations based solely on energy deposition in the stellar envelope are insufficient to account for the transition to the RG/RSG phase. Furthermore, we showed that the conventional “mirror principle”—which assumes that envelope expansion directly follows core contraction—is incomplete. Instead, we identified a refined form of the mirror principle governing post-main-sequence envelope evolution: the envelope expands or contracts in opposition to the motion of its inner boundary, located at radius $R_{\rm in}$ (denoted as $R_{\rm shell}$ in Paper~I). The values of $R_{\rm in}$ are initially set by helium-core contraction and are later regulated by nuclear energy generation in both the core and the shell. 

Paper~I also identified a distinct structural transition as stars approach the RG/RSG phase. During this transition, surface expansion temporarily stalls, while a substantial fraction of the interior mass is redistributed into the outer envelope. This redistribution increases the envelope opacity, producing an extended convective zone that penetrates toward the envelope base. In addition, in our previous studies \citep{Ou2023,Ou2025}, we found a bimodal distribution of supergiant radii, corresponding to red and blue supergiants, across stellar models spanning a wide range of masses and metallicities. However, the physical origin of the structural transition in RSG envelopes and the bimodal distribution of supergiant radii remain unclear. 

As a companion to Paper~I, which employed MESA stellar evolution models, in Paper~II we systematically explore the physically realizable configurations of stellar envelopes by constructing steady-state envelope models. This approach removes the constraints inherent in MESA stellar evolution models, which can compute envelope configurations only along particular evolutionary paths.

The structure of this paper is as follows. Section~\ref{sec:methods} formulates the physical problem and outlines the solution method. Section~\ref{sec:polytrope} introduces simplified polytropic envelope models, while Section~\ref{sec:realistic} presents more realistic models incorporating energy transport, tabulated opacities and equations of state, and convective mixing. Section~\ref{sec:discussions} discusses the strengths, limitations, and implications of our approach, and Section~\ref{sec:conclusion} concludes our findings.

\section{Methods} \label{sec:methods}

\begin{figure}[tbh]
\centering
\includegraphics[scale=0.8]{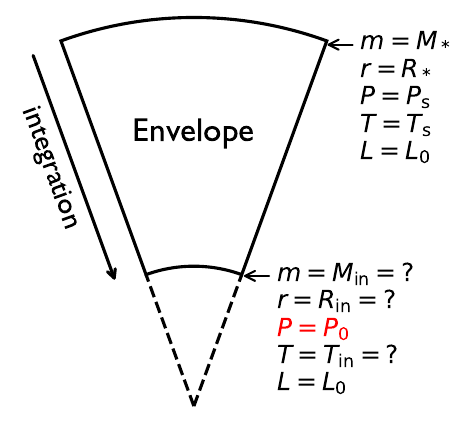}
\caption{Schematic diagram illustrating the problem setup in this study. We integrate the stellar structure equations inward from the surface down to the inner boundary of the envelope, which is defined by a termination pressure $P_0$.}
\label{fig:schematic}
\end{figure}

To examine the physically realizable configurations of stellar envelope, particularly the envelope's "response" to its inner boundary, we solve the stellar structure equations under boundary conditions motivated by the MESA models in Paper~I. As illustrated in Fig.~\ref{fig:schematic} our models compute only the envelope and do not include the core. The inner boundary of the envelope is not defined by a fixed enclosed mass, but by a fixed pressure $P = P_0$, corresponding to the physical conditions required to ignite the hydrogen burning of the shell. The outer boundary is located on the stellar surface, characterized by the stellar radius $R_*$ and the stellar mass $M_*$. Starting from the surface, we integrate the stellar equations inward until we reach $P = P_0$, thereby determining the $R_{\rm in}$ and $M_{\rm in}$. 

The first equation of steady-state stellar structure is mass conservation:

\begin{equation}
    \frac{dr}{dm} = \frac{1}{4\pi r^2 \rho},
\end{equation}
where $r$ is the radius and $\rho$ is the density. The second equation imposes hydrostatic equilibrium:
\begin{equation}
    \frac{dP}{dm} = -\frac{Gm}{4\pi r^4},
\end{equation}
where $P$ is the pressure and $G$ is the gravitational constant. Energy transport is described by the temperature ($T$) equation,
\begin{equation}
    \frac{dT}{dm} = -\frac{GmT}{4\pi r^4 P}\nabla,
\end{equation}
where the temperature gradient $\nabla$ depends on whether the energy is transported radiatively or convectively. 

The fourth stellar structure equation governs energy conservation,
\begin{equation}
\frac{dl}{dm} = \epsilon_{\rm nuc} + \epsilon_{\rm grav} - \epsilon_{\nu},
\end{equation}
where $l$ is the local luminosity, $\epsilon_{\rm nuc}$ is the nuclear energy generation rate, $\epsilon_{\nu}$ is the neutrino loss rate, and $\epsilon_{\rm grav}$ is the gravothermal term. In the envelope, both $\epsilon_{\rm nuc}$ and $\epsilon_{\nu}$ are negligible due to the low temperatures and densities. The remaining term, $\epsilon_{\rm grav}$, depends on time-dependent structural changes and therefore requires information from the evolutionary history. In our steady-state treatment, we assume a constant luminosity, $L_0$, throughout the envelope, which implies $\epsilon_{\rm grav} = 0$. The time-dependent energy conservation equation (Eq.~4) therefore need not be solved, leaving only Eqs.~(1)–(3) to define the steady-state envelope structure.

We obtain the envelope structure by numerically integrating Eqs.~(1)–(3), following the direct shooting method \citep[e.g.,][]{Kippenhahn2013}. For a given stellar radius $R_*$, surface boundary conditions are specified at $r = R_*$, including the surface temperature $T_{\rm s}$ and the surface pressure $P_{\rm s}$. The stellar structure equations are then integrated inward using the Runge–Kutta method RK23, implemented via the \texttt{solve\_ivp} function in \texttt{SciPy} \citep{Virtanen2020}, which solves the resulting system of ordinary differential equations as an initial-value problem. The calculation is completed when the pressure reaches a target value $P = P_0$, corresponding to the shell-envelope interface. Here, we adopt the fiducial value of $P_0 = 1.0\times 10^{16}\,{\rm dyne}\,{\rm cm}^{-2}$, estimated from Fig.~10 of Paper~I.

\subsection{Polytropic models}
 
 We construct a polytropic model, in which the gas obeys a power-law equation of state (EOS) relating the pressure $P$ to the density $\rho$:
\begin{equation}
    P= K\rho^{\gamma},
\end{equation}
where $K$ is the polytropic constant and $\gamma$ is the polytropic exponent. In this model, the pressure depends only on the density and is independent of the temperature. As a result, the equation of hydrostatic equilibrium is decoupled from energy transport \citep[e.g.,][]{Prialnik2009}. This simplification allows us to solve only the mass conservation and hydrostatic equilibrium equations (Eqs.~1 and 2), without explicitly computing the temperature structure. 

To integrate these equations, we impose a surface boundary condition on the pressure:
\begin{equation}
    P_{\rm s} = \frac{\tau_{\rm s} g}{\kappa_{\rm s}},
\end{equation}
where we adopt a surface optical depth of $\tau_{\rm s} = 2/3$. Because temperature is not included in the polytropic model, opacity tables that depend on temperature, density, and metallicity cannot be employed \citep[e.g.,][]{Iglesias1996}. We therefore assume a constant opacity $\kappa_{\rm s} = 0.34\,{\rm cm}^2\,{\rm g}^{-1}$, corresponding to electron scattering in fully ionized hydrogen. The surface gravity is given by $g = GM_*/R_*^2$, and hence the surface pressure scales as $P_{\rm s} \propto R_*^{-2}$.

\subsection{Realistic models}

We then construct more realistic models by solving Eqs.~(1)–(3), including energy transport, which require detailed microphysics such as the EOS and opacity. These data are taken from MESA (version 10108), with linear interpolation applied over the tabulated grids.

For the EOS, we adopt the MESA ``eosPT'' tables \citep{Paxton2011}, which take pressure and temperature as input parameters. We construct an effective EOS by averaging two tables with identical metallicity $Z = 0.02$ but hydrogen mass fractions $X = 0.6$ and $X = 0.8$, treating the result as representative of a composition with $X = 0.7$ and $Z = 0.02$. From this table, we extract the density $\rho (P,T)$, adiabatic temperature gradient $\nabla_{\rm ad} (P,T)$, 
specific heat at constant pressure $C_P (P,T)$, and the thermodynamic derivatives $\chi_T (P,T)\equiv (\partial\ln P/\partial\ln T)_\rho$ and $\chi_\rho(P,T) \equiv (\partial\ln P/\partial \ln \rho)_T$.

The opacity $\kappa (\rho,T)$ is taken from the MESA “OP$_{-}$gs98” table, which combines the OPAL \citep{Iglesias1993,Iglesias1996} and OP \citep{Seaton2005} opacities. For simplicity, we exclude Type~2 opacities that consider the enhancement of carbon and oxygen.

The surface boundary conditions consist of the surface temperature and pressure. The surface temperature ($T_{\rm s}$) is taken to be the effective temperature ($T_{\rm eff}$) of the star:
\begin{equation}
    T_{\rm s} = T_{\rm eff} = \bigg(\frac{L_0}{4\pi R_*^2\sigma} \bigg)^{1/4}.
\end{equation}
where $L_0$ is the stellar luminosity and $\sigma$ is the Stefan-Boltzmann constant. The luminosity is treated as a constant throughout the envelope and is specified as an input parameter. The surface pressure is determined using the ``simple photosphere'' prescription in MESA \citep{Paxton2011}:
\begin{equation}
    P_{\rm s} = \frac{\tau_{\rm s} g}{\kappa_{\rm s}}\bigg[1+1.6\times 10^{-4} \kappa_{\rm s}\bigg( \frac{L_0/L_{\odot}}{M_*/M_{\odot}} \bigg) \bigg],
\end{equation}
where we adopt $\tau_{\rm s} = 2/3$ and express the surface gravity as $g = GM_*/R_*^2$. Because $\kappa_{\rm s}$ depends on $\rho$ and $T$, and $\rho$ itself depends on $P$ through the EOS, this equation implicitly depends on $P_{\rm s}$. We therefore determine $P_{\rm s}$ numerically for a given $R_*$ using the \texttt{SciPy} function \texttt{root} with the ``hybr" method.

Unlike the polytropic models, the realistic models explicitly solve for the temperature profile via the energy transport equation. The mode of energy transport (radiation or convection) is determined by comparing the radiative temperature gradient $\nabla_{\rm rad}$ with the adiabatic gradient $\nabla_{\rm ad}$. The radiative gradient is given by
\begin{equation}
    \nabla_{\rm rad} = \frac{3}{16\pi acG} \frac{\kappa lP}{mT^4},
\end{equation}
where $a$ is the radiation density constant and $c$ is the speed of light, while $\nabla_{\rm ad}(P,T)$ is obtained from the EOS table. If the Schwarzschild criterion $\nabla_{\rm rad} < \nabla_{\rm ad}$ is satisfied, energy transport is purely radiative and the actual temperature gradient is set to $\nabla = \nabla_{\rm rad}$.

If $\nabla_{\rm rad} \geq \nabla_{\rm ad}$, convective transport is treated using mixing-length theory (MLT). Following the formulation of \citet{Kippenhahn2013}, the MLT framework solves for five quantities: the radiative flux $F_{\rm rad}$, the convective flux $F_{\rm con}$, the element temperature gradient $\nabla_{\rm e}$, the actual temperature gradient $\nabla$, and the convective velocity $v$. The total flux is the sum of the radiative and convective components:
\begin{equation}
    F_{\rm rad}+F_{\rm con} = \frac{4acG}{3}\frac{T^4 m}{\kappa P r^2} \nabla_{\rm rad}.
\end{equation}
The radiative flux is given by
\begin{equation}
    F_{\rm rad} = \frac{4acG}{3}\frac{T^4 m}{\kappa P r^2} \nabla.
\end{equation}
The convective velocity is calculated using
\begin{equation}
    v^2 = g\delta (\nabla-\nabla_{\rm e})\frac{l^2_{\rm m}}{8H_{P}},
\end{equation}
where $\delta \equiv - \left( \partial \ln \rho / \partial \ln T \right)_P = \chi_T / \chi_\rho$, $H_P \equiv -dr/d\ln P$ is the pressure scale height, and the mixing length is defined as $l_{\rm m} = \alpha H_P$ with $\alpha = 1.5$ in this work. The convective flux is then expressed as
\begin{equation}
    F_{\rm con} = \rho C_P T \sqrt{g\delta}\frac{l^2_{\rm m}}{4\sqrt{2}}H_P^{-3/2}(\nabla-\nabla_{\rm e})^{3/2}.
\end{equation}
The relationship between the four temperature gradients is given by
\begin{equation}
    \frac{\nabla_{\rm e}-\nabla_{\rm ad}}{\nabla-\nabla_{\rm e}} = \frac{6acT^3}{\kappa\rho^2 C_P l_{\rm m}v}.
\end{equation}

Directly solving the full set of five MLT equations is numerically challenging. We therefore adopt the formulation of \citet{Kippenhahn2013}, which introduces the dimensionless quantities
$U \equiv (3acT^3 / C_P \rho^2 \kappa l_{\rm m}^2)\sqrt{(8H_P / g\delta)}$ and
$W \equiv \nabla_{\rm rad} - \nabla_{\rm ad}$.
With this formulation, the actual temperature gradient can be obtained by solving a single cubic equation:
\begin{equation}
    (\xi-U)^3 + \frac{8U}{9} (\xi^2-U^2-W) = 0,
\end{equation}
where $\xi$ is defined by 
\begin{equation}
    \nabla = \xi^2+\nabla_{\rm ad}-U^2.
\end{equation}
We solve this equation for $\xi$ using the \texttt{SciPy} function \texttt{root\_scalar} with the "brentq" method, thus obtaining the actual temperature gradient $\nabla$, which is then used in the energy transport equation (Eq.~3).

\section{Results of Polytropic Models} \label{sec:polytrope}

To examine the relationship between $R_{\rm in}$ and $R_*$ in hydrostatic equilibrium alone, we construct polytropic models for a $25\,M_{\odot}$ star with different gas equations of state, characterized by various values of $\gamma$. We adopt a polytropic coefficient $K = 5.0 \times 10^{15}$ (in the appropriate cgs unit for each $\gamma$ value), to define the pressure-density relation (Eq.~5).

\begin{figure}[tbh]
\centering
\includegraphics[width=\columnwidth]{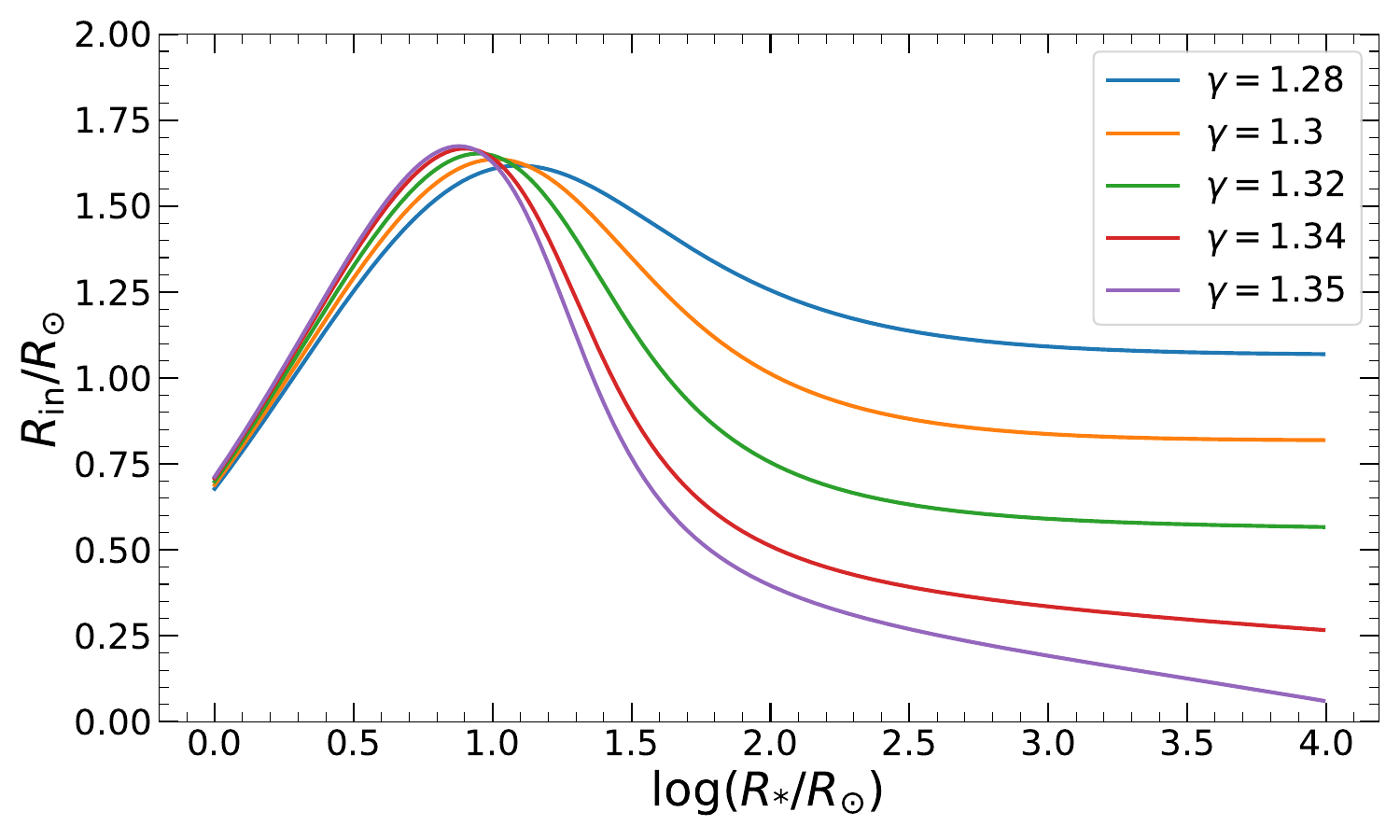}
\includegraphics[width=\columnwidth]{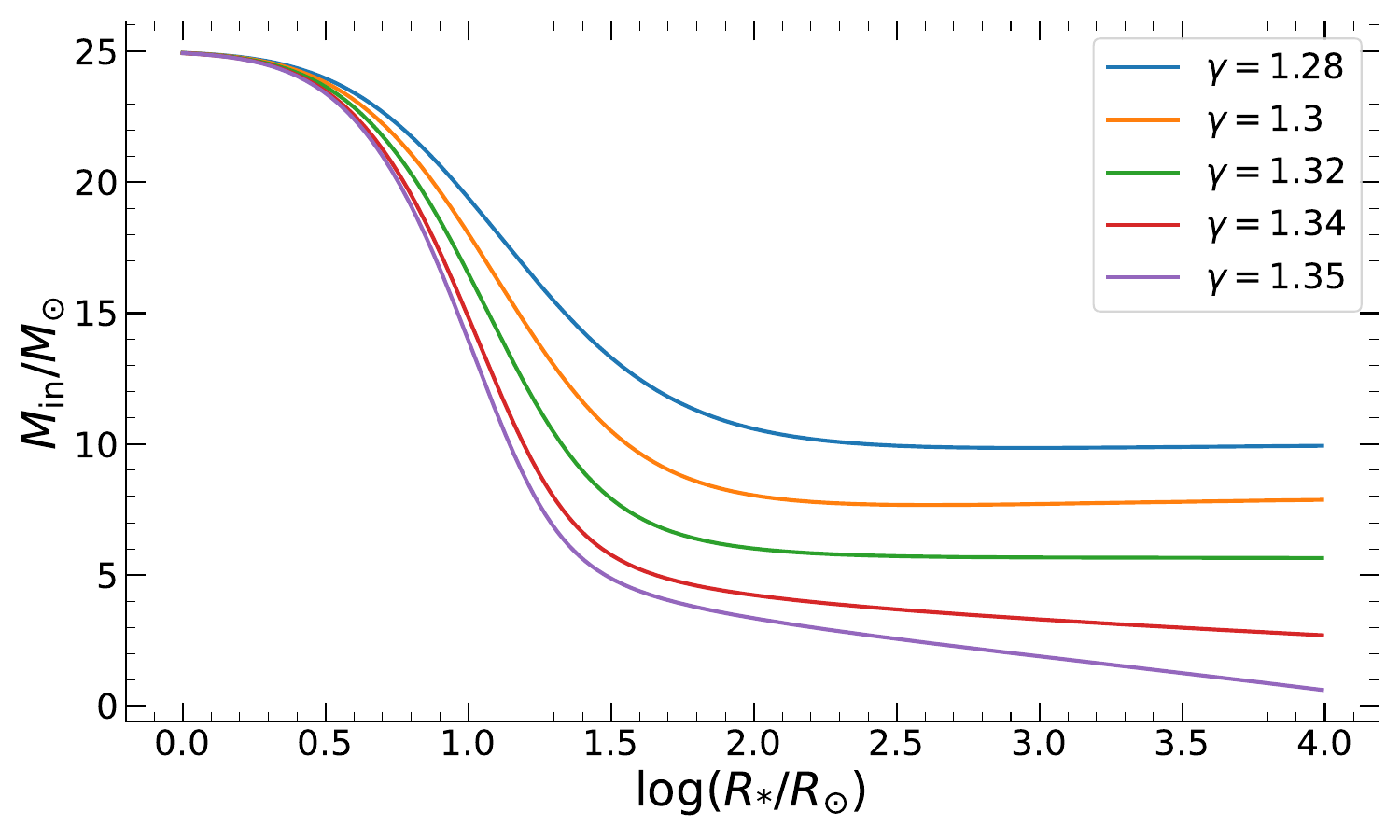}
\caption{Radius ($R_{\rm in}$, \textit{top panel}) and enclosed mass ($M_{\rm in}$, \textit{bottom panel}) at the envelope's inner boundary for polytropic models of $K=5.0\times 10^{15}$ (in the appropriate cgs unit for the corresponding $\gamma$) but different $\gamma$.}
\label{fig:gamma}
\end{figure}

Figure~\ref{fig:gamma} shows the resulting profiles of $R_{\rm in}$ and $M_{\rm in}$ as functions of $R_*$ for $\gamma = 1.28$–$1.35$. For $R_* \gtrsim 10\,R_{\odot}$, all these models consistently exhibit an anti-correlation between $R_*$ and $R_{\rm in}$, corresponding to the refined mirror principle identified Paper~I. At smaller radii ($R_* \lesssim 10\,R_{\odot}$), this anti-correlation no longer holds; however, the corresponding $M_{\rm in}$ values are far larger than those realized during post-main-sequence evolution (typically $M_{\rm in} \sim (7-10)\,M_{\odot}$ for $25\,M_{\odot}$ stars in the MESA models of Paper~1) and are therefore not physically relevant. This result demonstrates that hydrostatic equilibrium alone, without invoking energy transport, is sufficient to reproduce the refined mirror principle for physically realizable post-main-sequence envelope configurations.

\begin{figure}[tbh]
\centering
\includegraphics[width=\columnwidth]{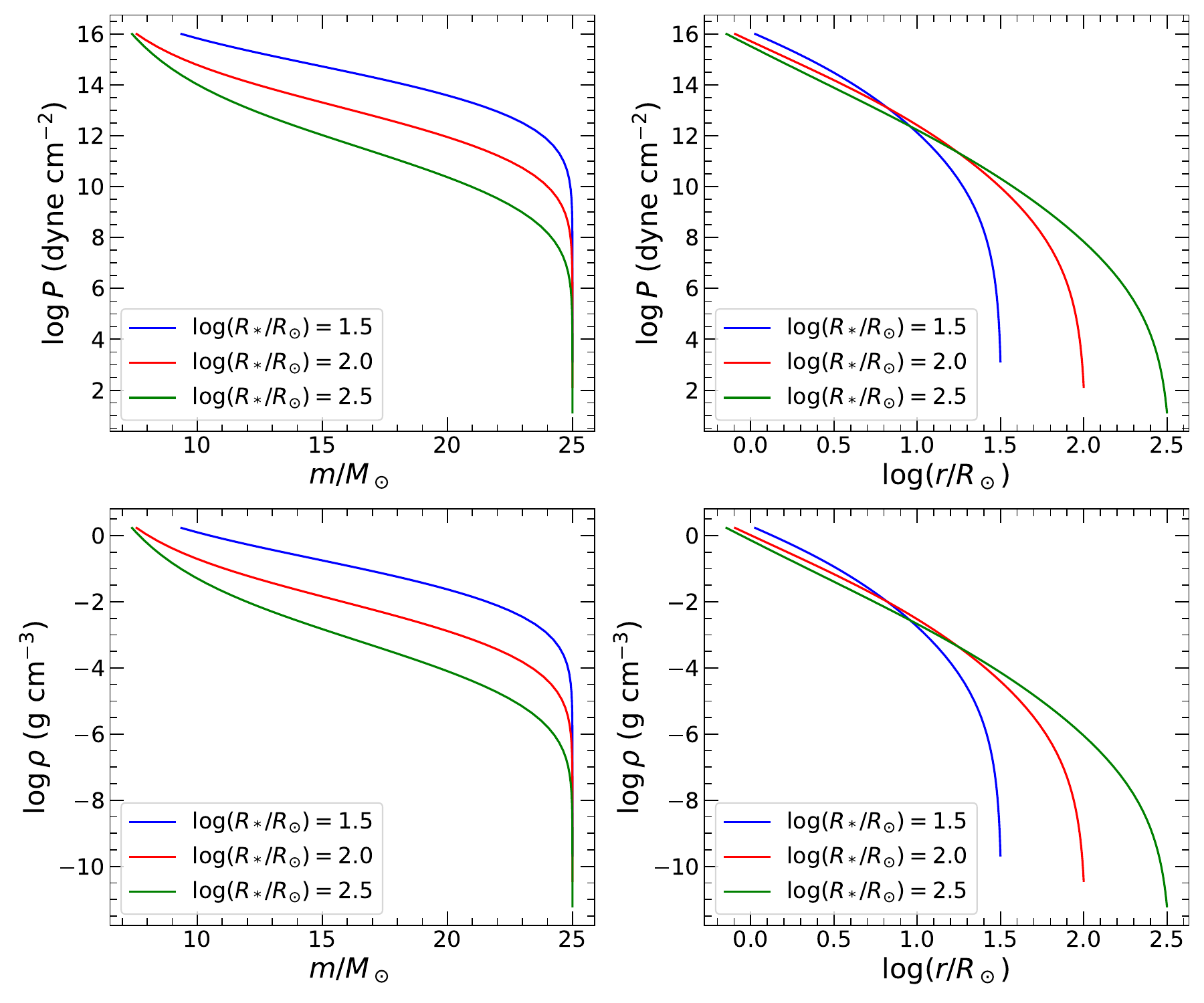}
\caption{Pressure and density profiles for polytropic models with $\gamma = 1.3$ and varying stellar radius $R_*$. Profiles are shown as functions of both mass and radius coordinates.}
\label{fig:logP_logRho_Rstar_gamma13}
\end{figure}

To illustrate the physical mechanism underlying the refined mirror principle, we plot the pressure and density profiles for the $\gamma = 1.3$ models of $\log(R_*/R_{\odot}) = 1.5$, 2.0, and 2.5 in Fig.~\ref{fig:logP_logRho_Rstar_gamma13}. We first examine the pressure gradients $|dP/dm|$ in the $\log P - m$ plot. The models of larger $R_*$ clearly exhibit larger $|dP/dm|$ close to the bottom of the envelope. According to the hydrostatic equilibrium equation (Eq.~2), the pressure gradient near the inner boundary scales as $|dP/dm|_{\rm in} \sim M_{\rm in}^2 R_{\rm in}^{-4}$. For the models considered here, the variations in $|dP/dm|_{\rm in}$ are dominated by the strong dependence on $R_{\rm in}^{-4}$, so a decrease in $R_{\rm in}$ leads to a rapid increase in $|dP/dm|_{\rm in}$. Because the pressure at the inner boundary is fixed at $P=P_0$, an increase in $|dP/dm|_{\rm in}$ can only be achieved by lowering the pressure throughout the envelope. Through the polytropic relation, this also implies a reduction in density. The resulting lower-density envelope occupies a larger volume, which naturally leads to an increase in stellar radius $R_*$.

These results are consistent with the MESA stellar evolution models presented in Paper~I. The physics of refined mirror principle can be summarized as follows. A smaller $R_{\rm in}$ implies stronger gravity at the shell-envelope interface, which demands a steeper pressure profile to maintain hydrostatic equilibrium. Because the pressure at the interface is fixed by the burning-shell conditions, this requirement is satisfied through envelope expansion, which reduces the pressure and density throughout the envelope. In this sense, the envelope “responds” to the inward movement of its inner boundary through a hydrostatic structural adjustment.


\section{Results of Realistic Models} \label{sec:realistic}

We next consider more realistic stellar envelope models that include temperature structure, tabulated equations of state and opacities, and convective energy transport treated with mixing-length theory (MLT). We present models with initial masses of 5 and $25\,M_{\odot}$, representing lower-mass and massive stars, respectively. For each model, we compute steady-state envelope structures assuming a fixed internal luminosity $L_0$, which is assumed to be constant throughout the envelope. Based on the MESA models presented in Paper I, we adopt $L_0 = (6$–$10)\times 10^{38}\,{\rm erg\,s}^{-1}$ for the $25\,M_{\odot}$ models and $L_0 = (4$–$12)\times 10^{36}\,{\rm erg\,s}^{-1}$ for the $5\,M_{\odot}$ models.

\begin{figure}[tbh]
\centering
\includegraphics[width=\columnwidth]{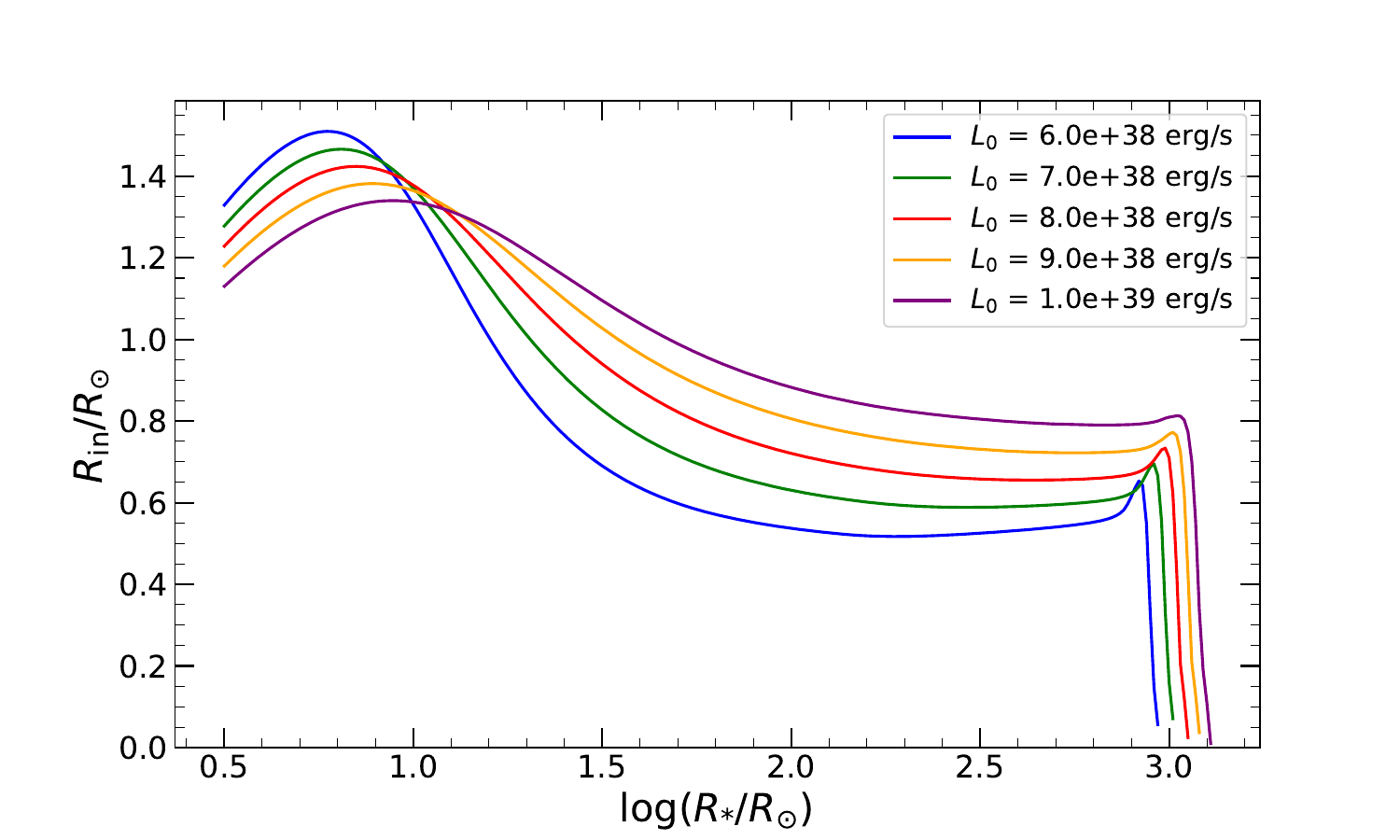}
\includegraphics[width=\columnwidth]{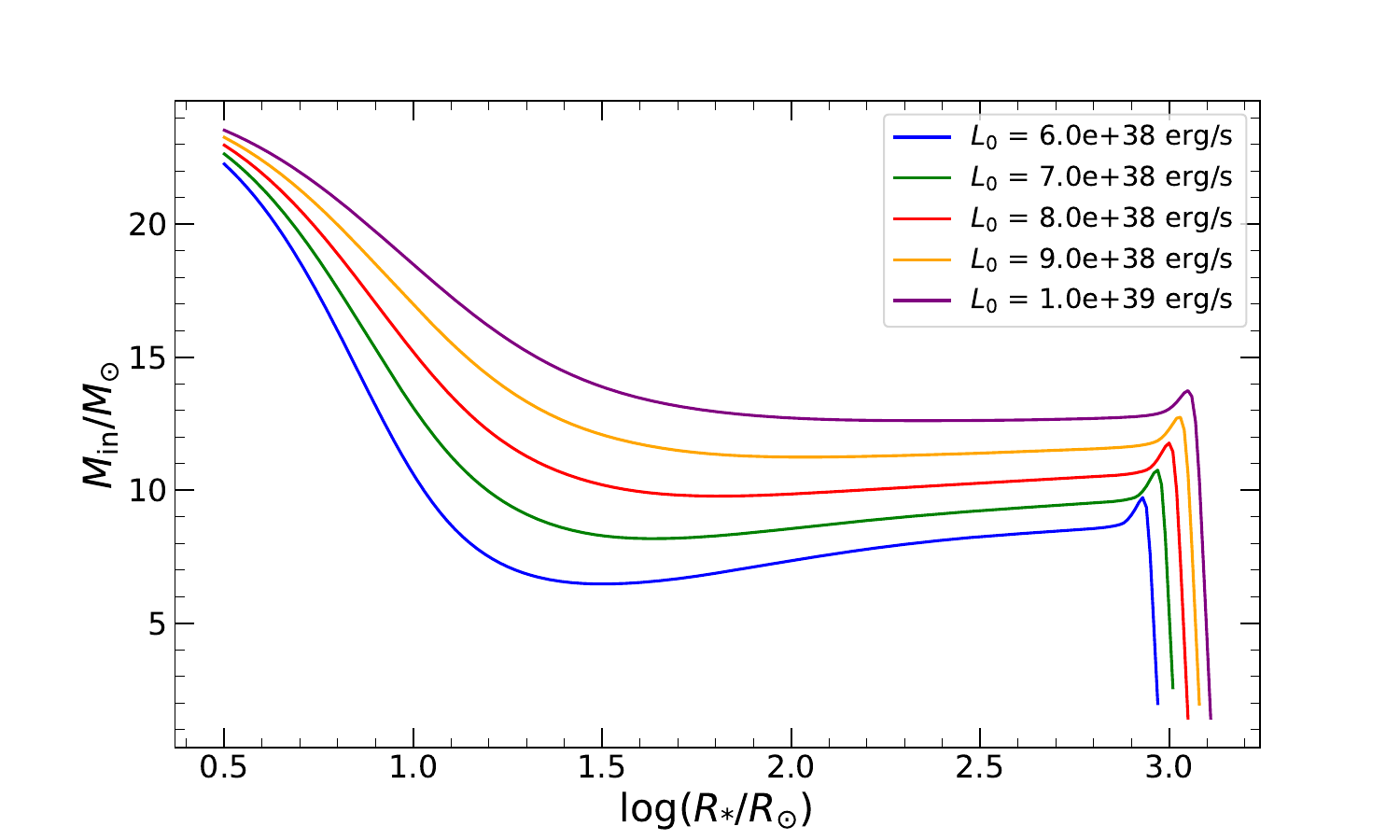}
\includegraphics[width=\columnwidth]{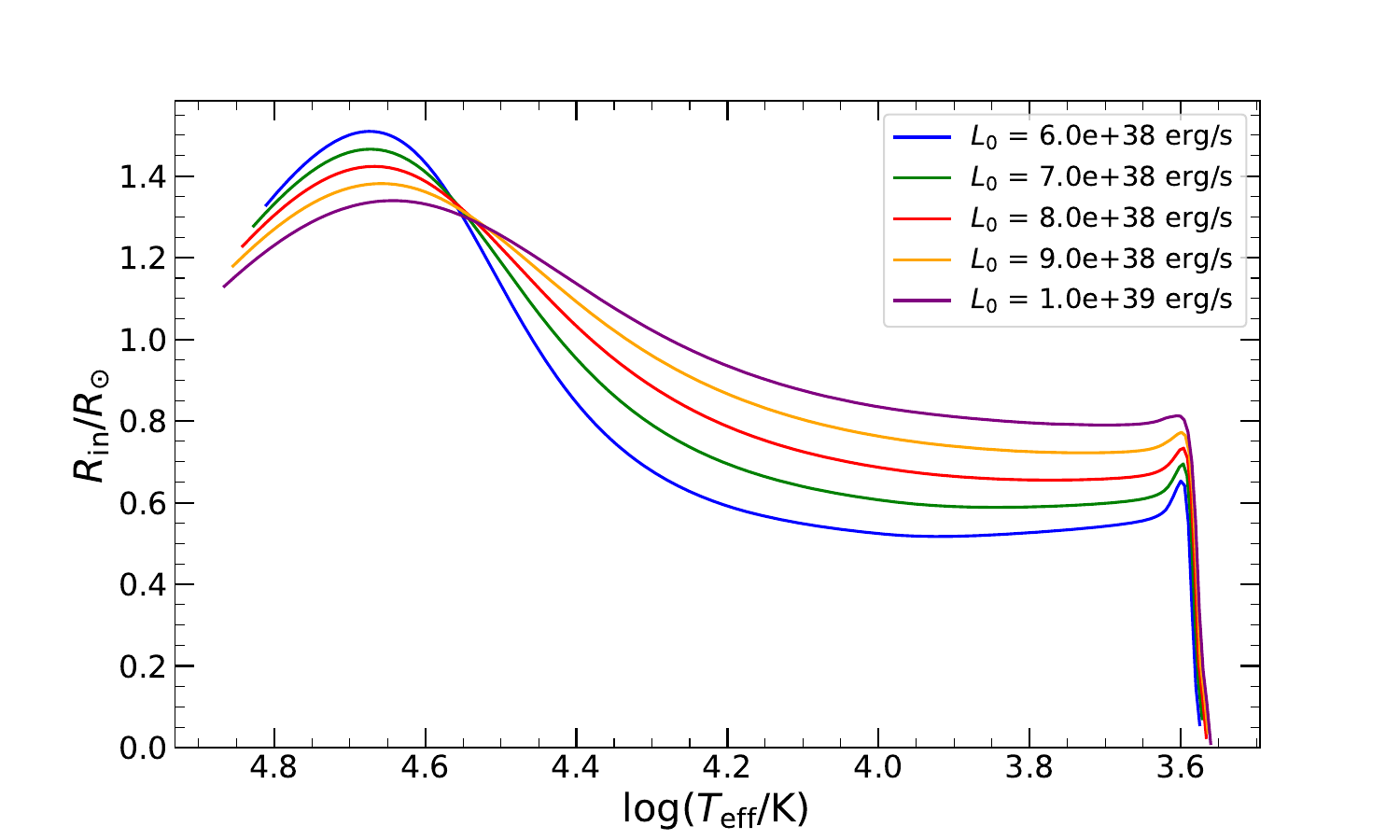}
\caption{$R_{\rm in}$ versus $R_*$ (\textit{top panel}), $M_{\rm in}$ versus $R_*$ (\textit{middle panel}), and $R_{\rm in}$ versus $T_{\rm eff}$ (\textit{bottom panel}) of $25\,M_{\odot}$ stars with different luminosities $L_0$.}
\label{fig:25Msun_MLT}
\end{figure}

\begin{figure}[tbh]
\centering
\includegraphics[width=\columnwidth]{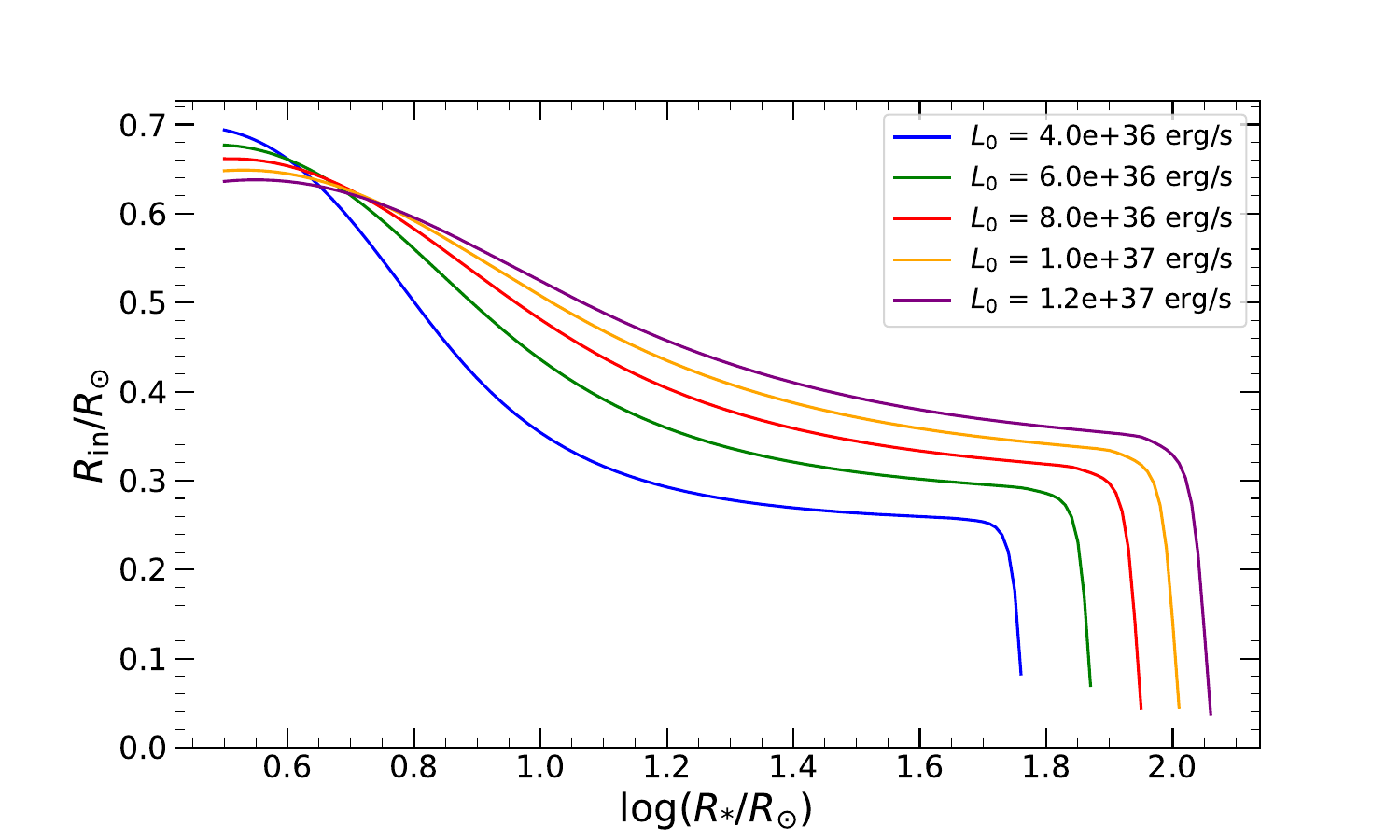}
\includegraphics[width=\columnwidth]{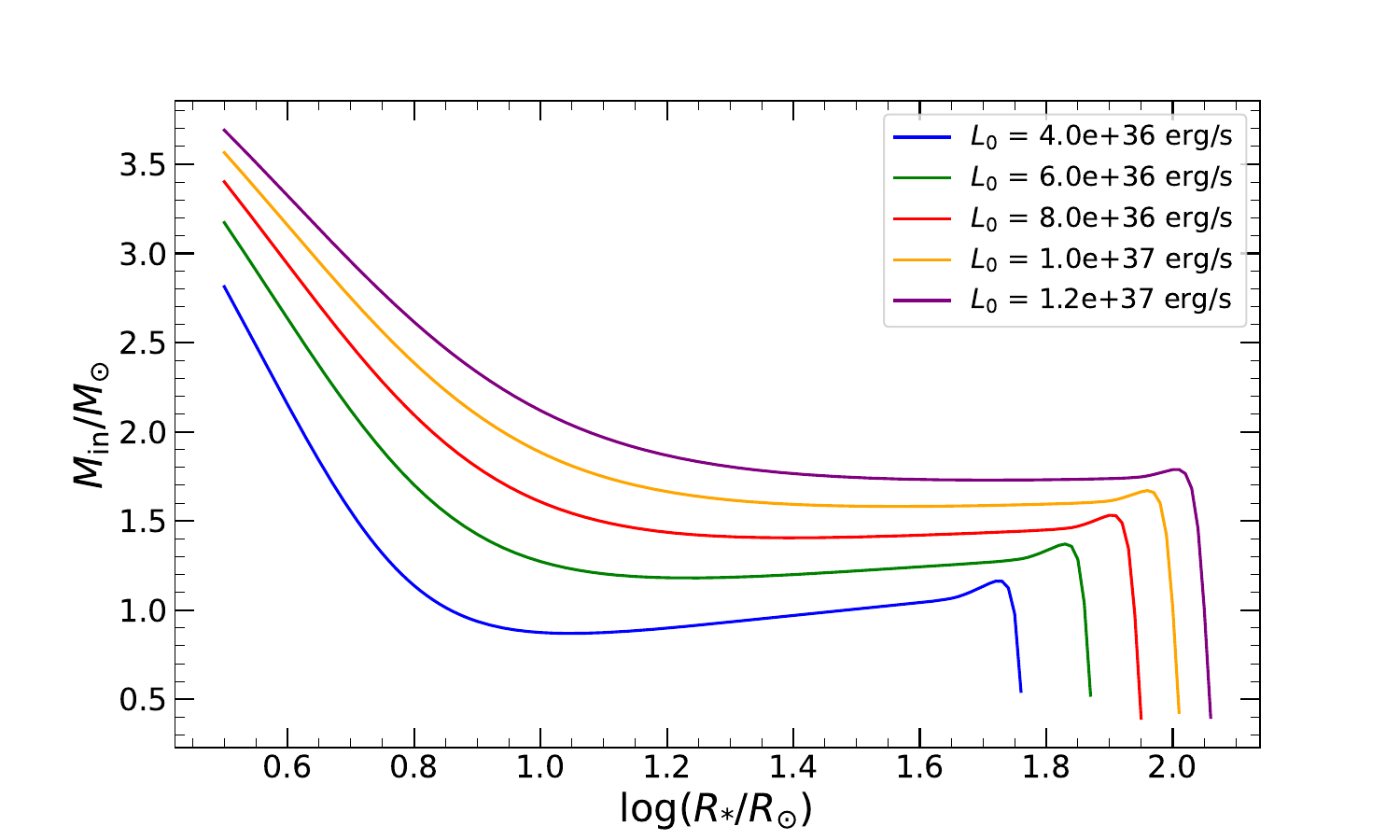}
\includegraphics[width=\columnwidth]{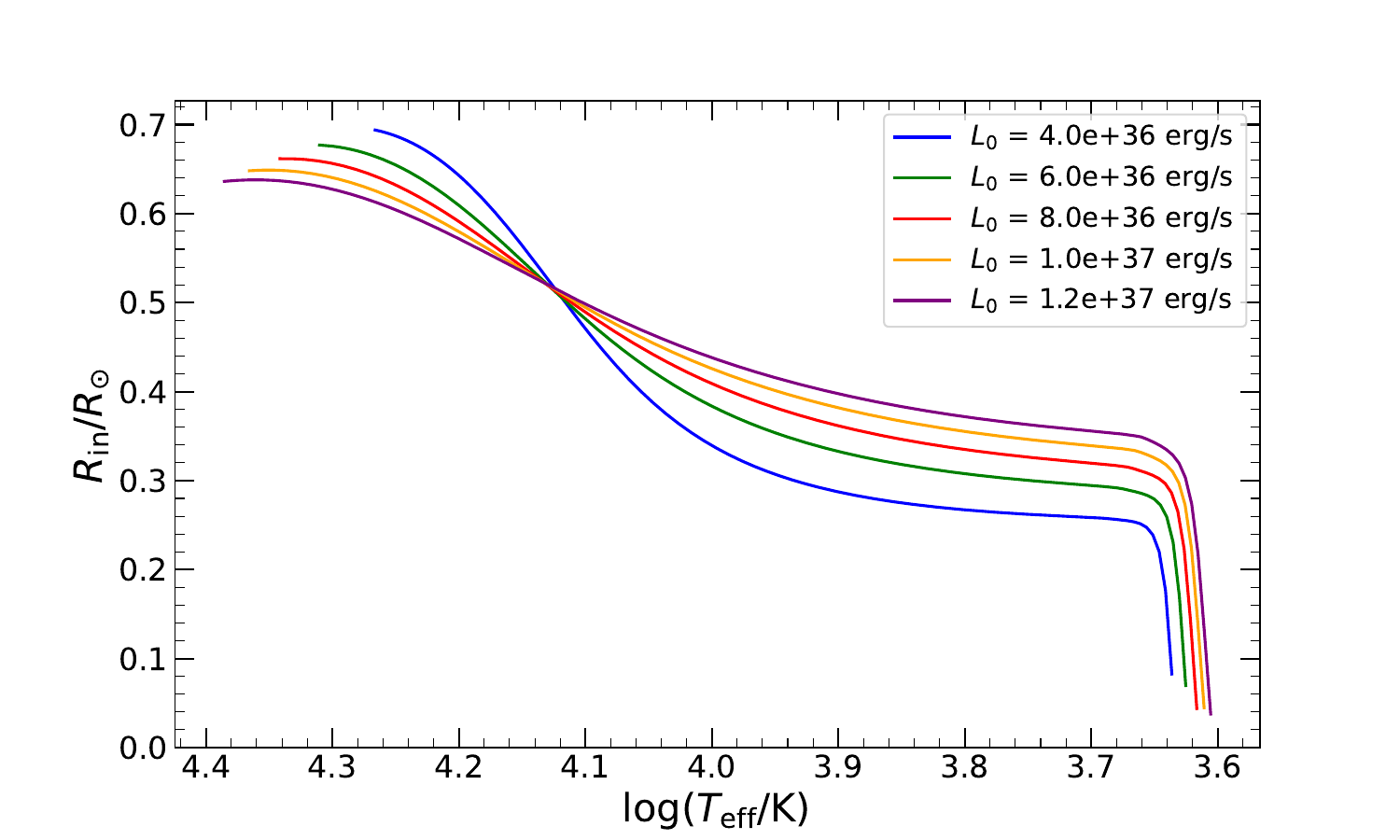}
\caption{$R_{\rm in}$ versus $R_*$ (\textit{top panel}), $M_{\rm in}$ versus $R_*$ (\textit{middle panel}), and $R_{\rm in}$ versus $T_{\rm eff}$ (\textit{bottom panel}) of $5\,M_{\odot}$ stars with different luminosities $L_0$.}
\label{fig:5Msun_MLT}
\end{figure}

Figs.~\ref{fig:25Msun_MLT} and \ref{fig:5Msun_MLT} show the results for the $25\,M_{\odot}$ and $5\,M_{\odot}$ models with different values of $L_0$, illustrating the relationships among $R_{\rm in}$, $M_{\rm in}$, $R_*$, and $T_{\rm eff}$. The overall behavior of $R_{\rm in}$ as a function of $R_*$ closely resembles that found in the polytropic models: the mirror principle holds when $M_{\rm in}$ lies within the physically realizable range for post-main-sequence stars, corresponding to $R_* \gtrsim 10\,R_{\odot}$ for $25\,M_{\odot}$ and $R_* \gtrsim$ a few $R_{\odot}$ for $5\,M_{\odot}$. 

A notable difference from the polytropic models emerges near the typical radii of RSGs ($R_* \sim 10^3\,R_{\odot}$) for $25\,M_{\odot}$ models and RGs ($R_* \sim 10^2\,R_{\odot}$) for $5\,M_{\odot}$ models. In this regime, $R_{\rm in}$ increases slightly before undergoing a sharp decline with increasing $R_*$. During this sharp decline, the effective temperatures for all models (both $25$ and $5\,M_{\odot}$) converge to $T_{\rm eff}\sim 4000\,{\rm K}$ (i.e., $\log(T_{\rm eff}/{\rm K}) \sim 3.6$), consistent with the characteristic surface temperatures of RGs and RSGs.

Importantly, the nearly vertical $R_{\rm in}$–$R_*$ relation near the RG/RSG radii effectively sets an upper limit to $R_*$. Even as $R_{\rm in}$ contracts to small values, the envelope cannot expand beyond this threshold. The RG/RSG phase therefore represents an upper limit for envelope expansion in steady-state solutions and behaves as an ultimate state, remaining largely insensitive to moderate changes in $R_{\rm in}$.


\subsection{Structural Transition Toward the Red Giant/Supergiant Phase}

\begin{figure*}[tbh]
\centering
\includegraphics[scale=0.4]{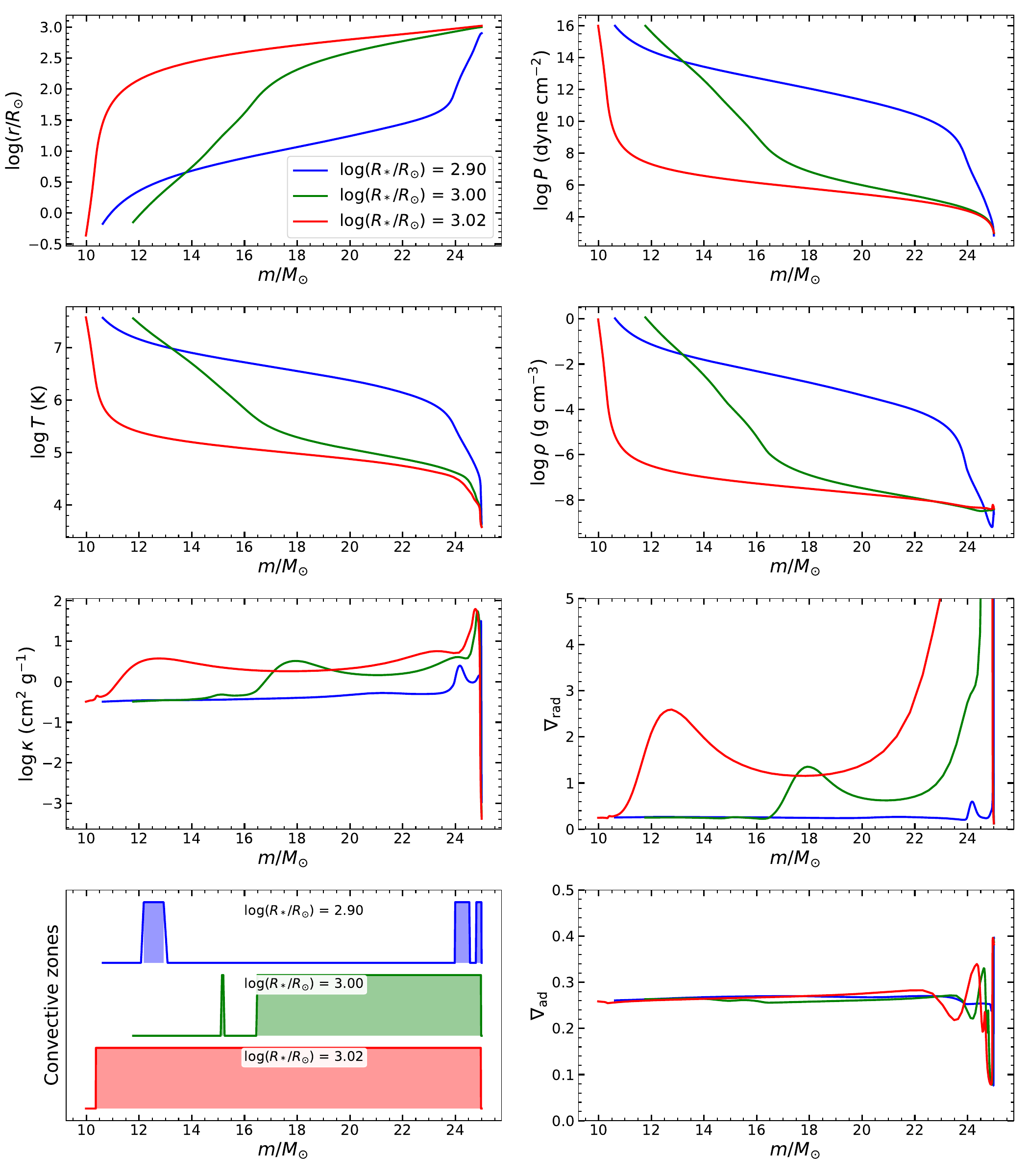}
\caption{Profiles of radius ($r$), pressure ($P$), temperature ($T$), density ($\rho$), opacity ($\kappa$), radiative temperature gradient ($\nabla_{\rm rad}$), convective zones, and adiabatic temperature gradient ($\nabla_{\rm ad}$) for $25\,M_{\odot}$ stars with different stellar radii. The blue, green, and red curves correspond to stellar radii of $\log (R_*/R_{\odot}) = 2.90$, 3.00, and 3.02 ($R_* \sim 800$, 1,000, and 1,050$\,R_{\odot}$), respectively, illustrating steady-state envelope solutions that roughly represent several time points as the star approaches the RSG phase. In the bottom-left panel, the color shade regions represent the distribution of convective zones.}
\label{fig:25Msun_MLT_profile}
\end{figure*}

To examine the stellar structures during the RSG formation, Fig.~\ref{fig:25Msun_MLT_profile} shows the structural profiles of a $25\,M_{\odot}$ star at three stages as the envelope expands toward the RSG regime, with stellar radii of $\log (R_*/R_{\odot}) = 2.9$, 3.0, and 3.02 (approximately $800\,R_{\odot}$, $1{,}000\,R_{\odot}$, and $1{,}050\,R_{\odot}$, respectively).

These profiles reveal pronounced structural changes during the final phase of expansion toward the RSG radii. Based on the radius plot in Fig.~\ref{fig:25Msun_MLT_profile}, for $R_* \sim 800\,R_{\odot}$, only a small fraction of the envelope mass occupies large radii $r > 100\,R_{\odot}$. The mass coordinate $m \sim 24\,M_{\odot}$ roughly corresponds to $r\sim 100\,R_{\odot}$, while only the outermost $\sim 1\,M_{\odot}$ of the envelope (from $m \sim 24-25\,M_{\odot}$) is distributed over the extended region between $r\sim 100-800\,R_{\odot}$. As $R_*$ increases from $800\,R_{\odot}$ to $1{,}050\,R_{\odot}$, the mass distribution changes dramatically, resulting in a qualitatively different radial mass distribution. At $R_* \sim 1{,}050\,R_{\odot}$, $m \sim 12\,M_{\odot}$ is redistributed to $r > 100\,R_{\odot}$.

This mass redistribution is also evident in the density profiles. At $R_* \sim 800\,R_{\odot}$, most of the envelope remains relatively compact, with densities $\rho > 10^{-4}\,{\rm g\,cm}^{-3}$, and only the outermost layers reach lower densities. In contrast, at $R_* \sim 1,050\,R_{\odot}$ the density drops sharply near the base of the envelope, decreasing from $\sim 1\,{\rm g\,cm}^{-3}$ at $m = 10\,M_{\odot}$ to $\lesssim 10^{-6}\,{\rm g\,cm}^{-3}$ at $m = 11\,M_{\odot}$.

\begin{figure}[tbh]
\centering
\includegraphics[scale=0.3]{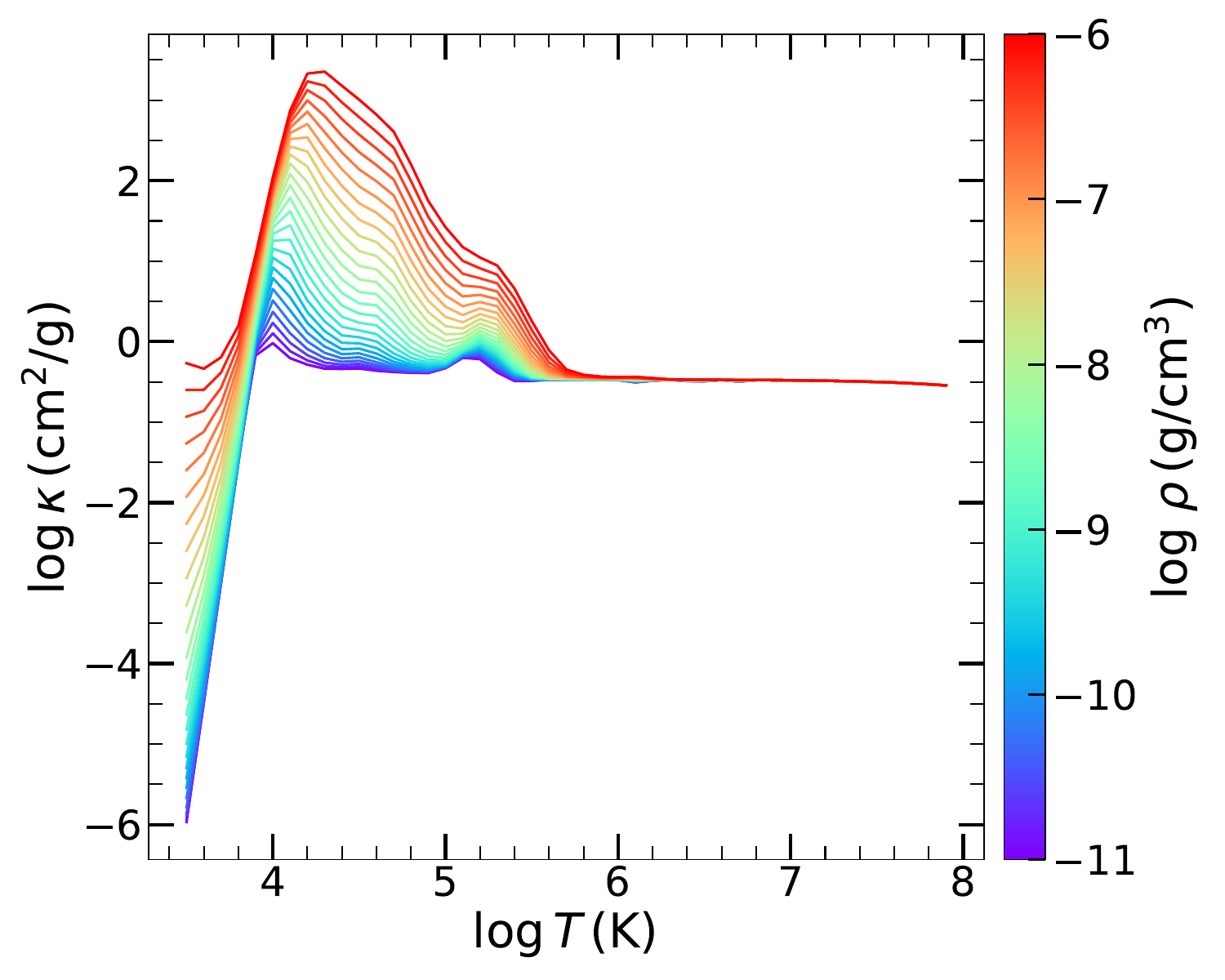}
\caption{Opacity as a function of temperature for various densities retrieved from the MESA "OP\_gs98" opacity tables at $X=0.7$ and $Z=0.02$.}
\label{fig:kappa}
\end{figure}

The temperature structure undergoes an equally dramatic transformation. At $R_* \sim 800\,R_{\odot}$, regions near $m \sim 23\,M_{\odot}$ remain hot, with temperatures close to $10^{6}\,{\rm K}$, and only the outermost layers cool below this value. By contrast, when the star reaches $R_* \sim 1{,}050\,R_{\odot}$, nearly the entire envelope has cooled to $T \lesssim 10^{6}\,{\rm K}$. This global cooling strongly affects the opacity. In the temperature range $10^{4}$–$10^{6}\,{\rm K}$, the opacity increases dramatically as the temperature decreases (Fig.~\ref{fig:kappa}). Across most of the envelope mass, the temperature declines by a factor of $\sim 30-100$ as $R_*$ increases from $\sim 800$ to $1{,}050\,R_{\odot}$, resulting in a substantial increase in opacity.

The resulting opacity enhancement drives a critical structural transition: the formation of an extended convective envelope. The increased opacity increases the radiative temperature gradient $\nabla_{\rm rad}$ well above the adiabatic gradient $\nabla_{\rm ad}$, triggering convection. As the star expands and progressively deeper layers cool into the high-opacity regime, the convective zone extends toward the base of the envelope. Eventually, nearly the entire envelope becomes convective. As shown in the bottom-left panel of Fig.~\ref{fig:25Msun_MLT_profile}, at $R_* \sim 800\,R_{\odot}$ only a few localized convective regions are present, each occupying a small fraction of the envelope mass. By $R_* \sim 1{,}000\,R_{\odot}$, the surface convective zone has extended to $m \sim 16.5\,M_{\odot}$, and by $R_* \sim 1{,}050\,R_{\odot}$ the convective region occupies nearly the entire envelope.

These structural changes—mass redistribution and the development of an extended convective envelope—are fully consistent with the stellar evolution models presented in Section~6 of Paper~I. Our steady-state solutions now reproduce this transition, demonstrating that the RG/RSG configuration is not merely a consequence of stellar evolution but a characteristic envelope structure. This result highlights the physical significance of the structural transition. The nearly vertical decline of $R_{\rm in}$ with respect to $R_*$ near $R_* \sim 10^3\,R_{\odot}$ coincides with the completion of the phase transition to the RG/RSG regime, indicating that further radial expansion is no longer possible. After this transition, the pressure and temperature gradients throughout the envelope become extremely shallow. Any further increase in radius would flatten these gradients even more, preventing the inward integration from reaching the termination pressure $P_0$ within the available stellar mass. Consequently, no steady-state envelope solution can exist for stars beyond this limiting radius.

\subsection{Physical Origin of the Red Giant/Supergiant Solution}

\begin{figure}[tbh]
\centering
\includegraphics[scale=0.5]{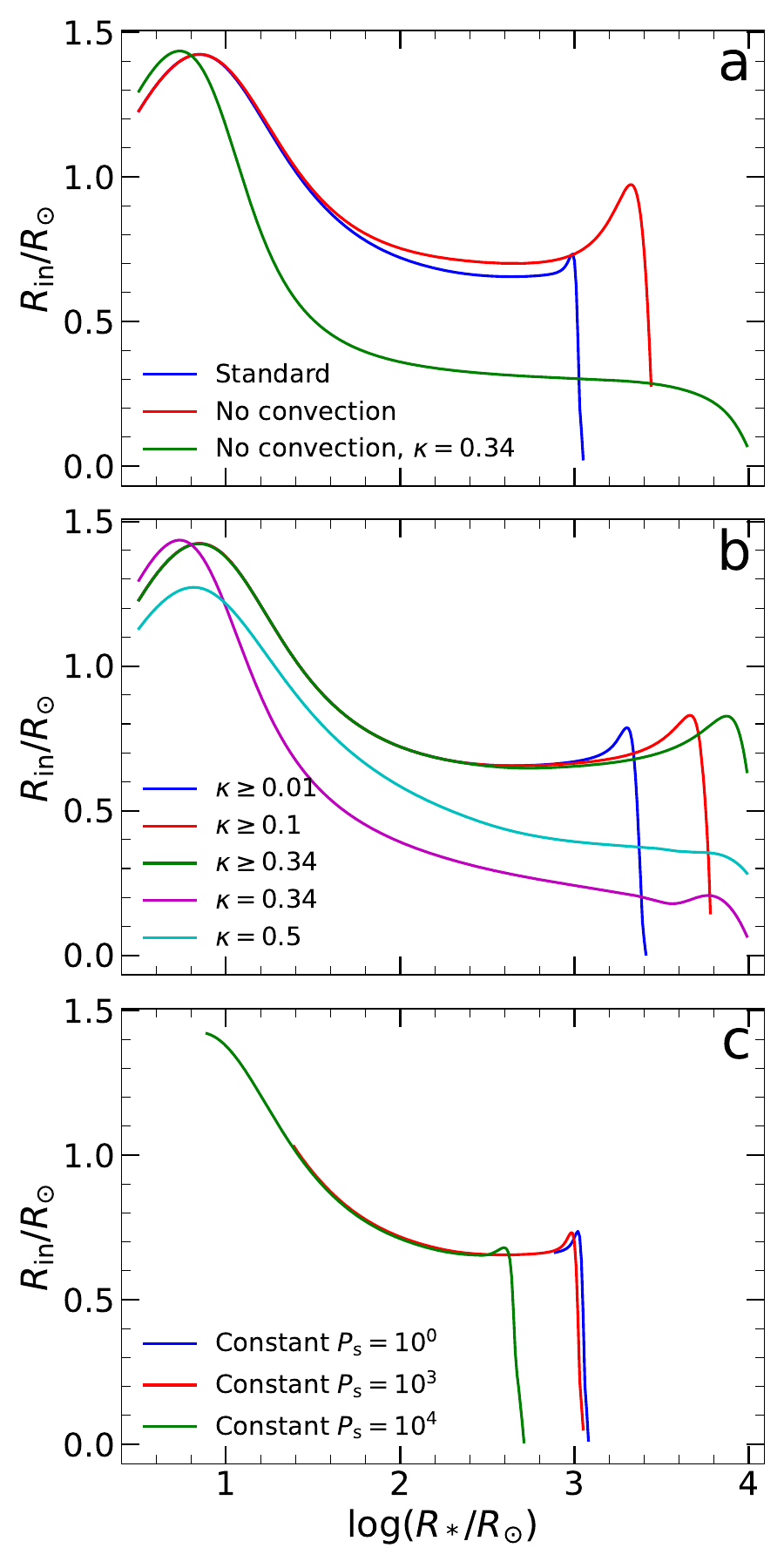}
\caption{Effects of convection, opacity, and surface pressure on $R_{\rm in}$–$R_*$. All models assume a stellar mass of $25\,M_{\odot}$ and a fixed luminosity of $L_0 = 8.0 \times 10^{38}\,{\rm erg\,s}^{-1}$. Panel (a): standard model, no-convection model, and no-convection model with a constant opacity of $0.34\,{\rm cm}^2\,{\rm g}^{-1}$ (electron scattering). Panel (b): models with imposed opacity floors of $\kappa \geq 0.01$, 0.1, and $0.34\,{\rm cm}^2\,{\rm g}^{-1}$, as well as models with constant opacities $\kappa =0.34$, and $0.5\,{\rm cm}^2\,{\rm g}^{-1}$. Panel (c): models with fixed surface pressures of $10^0$, $10^3$, and $10^4\,{\rm dyne\,cm}^{-2}$.}
\label{fig:25Msun_4cases}
\end{figure}

To evaluate the impact of stellar parameters on the RG/RSG solutions, we conduct a series of controlled experiments by varying key physical ingredients, including the treatment of convection, opacity interpolation from tables, and surface boundary conditions from atmospheric models. All models assume a stellar mass of $25\,M_{\odot}$ and a fixed luminosity of $L_0 = 8.0 \times 10^{38}\,{\rm erg\,s}^{-1}$. Figure~\ref{fig:25Msun_4cases} shows the resulting $R_{\rm in}$–$R_*$ relations for these models; the model setups in Panels (a)–(c) are listed below.

\begin{enumerate}
\item[(a)-1] Standard model: includes MLT convection, tabulated opacities, and a simple atmospheric prescription for the surface pressure boundary condition.
\item[(a)-2] No convection: MLT convection is disabled and energy transport is purely radiative, with $\nabla=\nabla_{\rm rad}$ throughout the envelope.
\item[(a)-3] No convection and constant opacity: MLT convection is disabled and a constant opacity of $\kappa = 0.34\,{\rm cm}^2\,{\rm g}^{-1}$ (electron scattering in fully ionized hydrogen) is adopted throughout the envelope.
\item[(b)-1] Constant opacity: a uniform opacity of $\kappa = 0.34$ or $0.5\,{\rm cm}^2\,{\rm g}^{-1}$ is imposed throughout the envelope.
\item[(b)-2] Opacity floor: a minimum opacity is enforced, with $\kappa \ge 0.01$, $0.1$, $0.34$, and $1.0\,{\rm cm}^2\,{\rm g}^{-1}$.
\item[(c)] Constant surface pressure: the outer boundary condition is replaced by a fixed surface pressure, with $P_{\rm s}=10^{4}$, $10^{3}$, and $10^{0}\,{\rm dyne\,cm}^{-2}$.
\end{enumerate}

We first compare the standard model with the cases in which convection is suppressed. The RSG solution—characterized by the abrupt convergence toward a limiting $R_*$—persists even when convection is disabled, although it is shifted to a larger radius, $R_* \sim 1.5\times10^3\,R_\odot$. However, when the tabulated opacity is further removed and a constant electron-scattering opacity of $\kappa=0.34\,{\rm cm}^2\,{\rm g}^{-1}$ is adopted, the RSG limiting radius disappears altogether. This shows that opacity plays a crucial role in the formation of RSGs.

We further test the role of opacity in Fig.~\ref{fig:25Msun_4cases}b using models that retain convection. With constant opacity, the abrupt convergence to an RSG solution does not occur, whereas imposing a minimum opacity floor on tabulated opacities shifts the RSG solution to progressively larger radii: $R_* \sim 1,500\,R_{\odot}$ for $\kappa \geq 0.01\,\rm{cm}^2\,\rm{g}^{-1}$, $R_* \sim 5,000$–$6,000\,R_{\odot}$ for $\kappa \geq 0.1\,\rm{cm}^2\,\rm{g}^{-1}$, and even $R_* \sim 10,000\,R_{\odot}$ for $\kappa \geq 0.34\,\rm{cm}^2\,\rm{g}^{-1}$. 

We now examine how opacity sets the limiting radius of the RSG solution. In our models, opacity influences the stellar structure in two primary ways, through Eqs.~(8) and (9). The first operates via the surface boundary condition. From Eq.~(8), the surface pressure $P_{\rm s}$ depends on the surface opacity $\kappa_{\rm s}$, with higher opacities leading to lower $P_{\rm s}$. However, the experiments shown in Fig.~\ref{fig:25Msun_4cases}c demonstrate that variations in $P_{\rm s}$ alone cannot shift the RSG solution to substantially larger radii and cannot cause the solution to vanish. While setting surface pressure at $P_{\rm s}=10^4\,{\rm dyne}\,{\rm cm}^{-2}$ reduces the limiting $R_*$, lowering $P_{\rm s}$ to $10^3\,{\rm dyne}\,{\rm cm}^{-2}$— or even to $10^0\,{\rm dyne}\,{\rm cm}^{-2}$— produces only a marginal increase in the maximum radius of the RSG. Therefore, changes in the surface boundary condition alone cannot explain the essential role of opacity in shaping the RSG solution (Fig.~\ref{fig:25Msun_4cases}b).

The second role of opacity is its influence on radiative energy transport. In the radiative regime, the temperature gradient obeys
\begin{equation}
    \frac{dT}{dm}= -\frac{3}{64\pi^2 ac}\frac{\kappa l}{r^4 T^3}.
\end{equation}
In our models, the luminosity is fixed to $l=L_0$ throughout the envelope, and the surface condition gives $L_0 \sim R_*^2 T_{\rm eff}^4$. Near the stellar surface, the temperature gradient scales approximately as
\begin{equation}
    \bigg|\frac{dT}{dm}\bigg|_{\rm s} \sim \kappa_{\rm s} T_{\rm eff}^{5}.
\end{equation}
In the H$^{-}$ opacity regime, $\kappa_{\rm s}$ is highly sensitive to temperature. According to the MESA "OP$_{-}$gs98" opacity table (Fig.~\ref{fig:kappa}), at $\rho \sim 10^{-8}\,{\rm g\,cm}^{-3}$ the opacity scales approximately as $\kappa \sim T^{11}$ over the range $3,000$–$10,000\,{\rm K}$, implying $|dT/dm|_{\rm s}\sim T_{\rm eff}^{16}$, highlighting the extreme temperature sensitivity of radiative energy transport for the physical condition near the stellar surface. As the star expands and $T_{\rm eff}$ declines, the temperature gradient therefore becomes increasingly shallow, driving the structural transition to the RG/RSG configuration. In the standard model shown in Fig.~\ref{fig:25Msun_MLT_profile}, when $\log(R_*/R_{\odot}) = 3.02$ (i.e., $R_* \sim 1,050\,R_{\odot}$), the surface opacity falls below $10^{-3}\,{\rm cm}^2\,{\rm g}^{-1}$, producing a markedly flatter temperature profile than in models with smaller $R_*$. 

\begin{figure*}[tbh]
\centering
\includegraphics[scale=0.4]{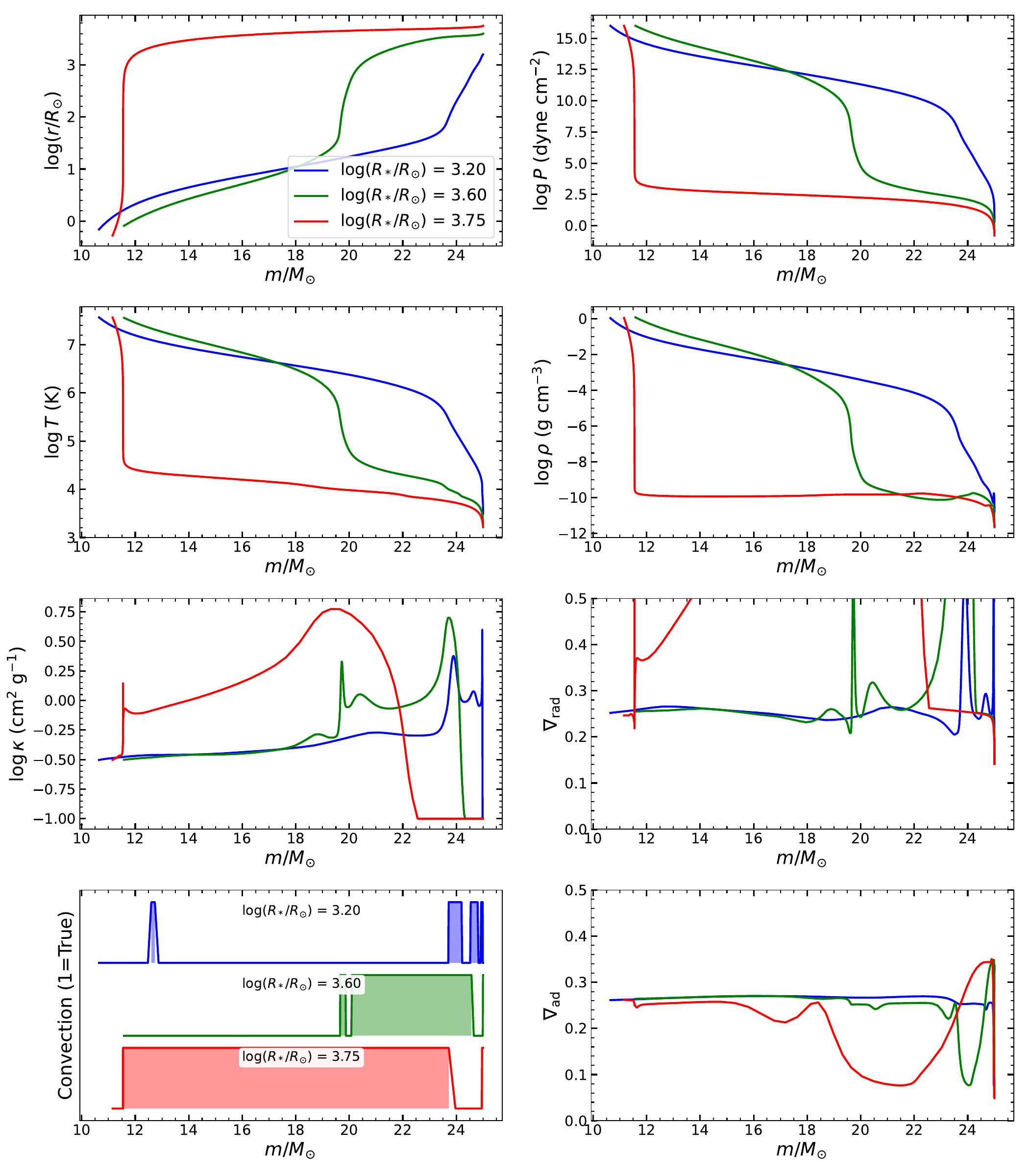}
\caption{Profiles for $25\,M_{\odot}$ stars, similar to those in Fig.~\ref{fig:25Msun_MLT_profile}, but with a minimum opacity floor of $\kappa \geq 0.1\,{\rm cm}^{2}\,{\rm g}^{-1}$. The blue, green, and red curves correspond to stellar radii of $\log (R_*/R_{\odot}) = 3.20$, 3.60, and 3.75, respectively, illustrating steady-state envelope solutions that roughly represent several time points as the star approaches the RSG phase.  In the bottom-left panel, the color shade regions represent the distribution of convective zones.}
\label{fig:kapmin_profile}
\end{figure*}

If an opacity floor is imposed to prevent $\kappa_{\rm s}$ from becoming too small, the surface temperature gradient $|dT/dm|_{\rm s}$ is no longer flattened at the same $R_*$. This is because, at a given $R_*$, $T_{\rm eff}$ is fixed, so $|dT/dm|_{\rm s}$ is controlled primarily by $\kappa_{\rm s}$. As shown in Fig.~\ref{fig:kapmin_profile}, imposing an opacity floor of $\kappa \geq 0.1\,{\rm cm}^2\,{\rm g}^{-1}$ prevents the surface layers from developing a flattened $|dT/dm|$ even when the stellar radius increases to $\log(R_*/R_{\odot}) = 3.2$. Only when the star expands further to $\log (R_*/R_{\odot})=3.75$ does a substantial fraction of the envelope enter the H$^{-}$ opacity regime and reaches the imposed floor value of $\kappa= 0.1\,{\rm cm}^2\,{\rm g}^{-1}$. At this point, the overall opacity becomes low enough to flatten the temperature profile and trigger the structural transition to the RSG configuration.

In summary, the temperature profile of the stellar envelope is highly sensitive to its opacity, particularly in the low-temperature regime of $T < 10{,}000$ K where H$^{-}$ opacity dominates. The opacity-driven flattening of $|dT/dm|$ is the key mechanism underlying the structural transition to the RG/RSG phase. In this framework, the RSG solution is determined by the low opacity of H$^{-}$ near the surface, which does not exist when a constant electron-scattering opacity is adopted.

This maximum radius of envelope expansion corresponds to the classical Hayashi limit, or Hayashi line \citep{Hayashi1961}, a nearly vertical locus in the HR diagram along which stars attain similar effective temperatures over a wide range of luminosities. It represents hydrostatic solutions for fully convective stellar envelopes, while the region to the right of this line in Hertzsprung-Russell (HR) diagram is forbidden, as no hydrostatic solutions exist within the framework of MLT. As demonstrated by our results, once the envelope becomes fully convective following the structural transition, the stellar structure settles onto this limiting configuration, beyond which further radial expansion is not possible. Accordingly, our models with different values of $L_0$ converge to a limiting effective temperature of $\sim 4,000\,{\rm K}$ (Fig.~\ref{fig:25Msun_MLT} and \ref{fig:5Msun_MLT}), consistent with the near-vertical nature of the Hayashi line.

The Hayashi limit arises from the intersection of radiative atmospheric solutions with the adiabatic $P$–$T$ stratification of a fully convective stellar interior, and is therefore highly sensitive to the envelope opacity \citep{Hayashi1961,Kippenhahn2013}. In our steady-state models, artificial modifications to the opacity produce substantial changes in the maximum radius of the convective envelope, which can be interpreted as shifts of the Hayashi line. This strong dependence of the maximum radius on opacity is consistent with the classical picture of the Hayashi limit.


\subsection{Red and Blue Solutions}

Our steady-state envelope solutions provide a physical framework for understanding the distribution of giant and supergiant stars in the HR diagram. Based on the $R_{\rm in}$–$R_*$ and $R_{\rm in}$–$T_{\rm eff}$ relations shown in Fig.~\ref{fig:25Msun_MLT} and \ref{fig:5Msun_MLT}, the solutions can be broadly divided into three distinct regimes.

\paragraph{(1) Red giant/supergiant solution.} Near $T_{\rm eff} \sim 4,000\,{\rm K}$, the RG/RSG solution represents a limiting configuration for envelope expansion. It corresponds to a distinct structural state, fundamentally different from that of more compact stars. As a star approaches this regime, it undergoes a pronounced structural transition characterized by a strongly flattened temperature gradient, substantial outward redistribution of the envelope mass, and the development of an extended convective envelope.

\paragraph{(2) Blue giant/supergiant regime.} At higher effective temperatures, $T_{\rm eff} \sim 10,000$–$30,000\,{\rm K}$, a wide range of steady-state envelope solutions exist. For a given luminosity, a smaller $R_{\rm in}$ generally corresponds to a larger $R_*$, resulting in a continuous range of blue giant and supergiant configurations.

\paragraph{(3) Yellow gap.} Between $T_{\rm eff} \sim 4,000$–$10,000\,{\rm K}$, the $R_{\rm in}$ curves are nearly horizontal with respect to $R_*$ or $T_{\rm eff}$, indicating that even a modest inward shift of $R_{\rm in}$ leads to a large change in the stellar radius, implying a rapid evolution across this temperature range. This behavior naturally explains the observed Hertzsprung gap, where stars are rarely found because they traverse this region quickly. Consequently, giant and supergiant stars tend to converge either to the red or blue branches, rather than remaining in the intermediate yellow regime.

\section{Discussions} \label{sec:discussions}

Compared to full stellar evolution calculations, the principal advantage of our steady-state approach is that it allows systematic exploration of stellar envelope structures across a wide range of physical parameter combinations, rather than being restricted to specific evolutionary tracks. This flexibility enables us to isolate the problem from evolutionary effects and to identify the fundamental mechanisms that govern the envelope structure. In addition, modifications to physical inputs, such as opacity, surface boundary conditions, or convective prescriptions, often cause stellar evolution calculations to stall or fail. By contrast, steady-state envelope models are independent of prior evolutionary history and therefore provide a practical framework for testing various physical assumptions.

Another key feature of our approach is that we solve only the envelope structure, with the inner boundary defined by a fixed pressure condition. This treatment allows us to focus directly on envelope physics and, in particular, to examine the refined mirror principle within the physically relevant range of $M_{\rm in}$. By excluding the nuclear-burning core, we also avoid the additional complexity introduced by nuclear energy generation and composition evolution, which are not essential for understanding the envelope response investigated here.

We also acknowledge several important caveats of our method. First, we assume a constant luminosity throughout the envelope, which allows us to avoid the gravothermal energy term, $\epsilon_{\rm grav}$, and thereby removes explicit time dependence from the problem. As shown in Section~\ref{sec:realistic}, the envelope structure is sensitive to the adopted luminosity, yet our steady-state models cannot determine which luminosities are realized during stellar evolution. In reality, luminosity evolves significantly during the post-main-sequence phase. As demonstrated in Paper~I, the luminosity at the base of the envelope correlates strongly with the local radius, $R_{\rm in}$. Consequently, a star does not evolve along a single $R_{\rm in}$–$R_*$ curve at fixed luminosity (Figs.~\ref{fig:25Msun_MLT} and \ref{fig:5Msun_MLT}), but instead traverses multiple such curves as its luminosity changes. The relations derived from our constant-luminosity models should therefore be interpreted as generic structural trends rather than as evolutionary paths. In this sense, the steady-state approach sacrifices information about which configurations are realized during actual stellar evolution.

Second, because our calculations isolate the envelope and omit the core, they do not capture structural constraints imposed by the core properties. To partially mitigate this limitation, we use MESA stellar evolution models to guide the choice of inner boundary conditions and to assess the physical relevance of the resulting envelope solutions. Nevertheless, by decoupling the envelope from the core, we inevitably lose information about the physical constraints imposed by the core structure.

Despite these limitations, our steady-state framework provides a powerful tool for probing the fundamental physics of stellar envelopes. By solving the envelope structure independently of stellar evolutionary constraints, this approach yields a clear physical insight into the origin of RG/RSG solutions. Beyond the applications presented here, the method may be extended to other evolutionary phases, including advanced stages following core carbon burning.

\section{Conclusions}\label{sec:conclusion}

To investigate the physical origin of envelope expansion toward the RG/RSG phase, we constructed steady-state stellar envelope models and solved for their internal structures. This approach allows us to explore a wide range of physically realizable envelope configurations and leads to the following key conclusions.

\paragraph{(1) The refined mirror principle for envelope expansion.} Even simplified polytropic models without energy transport reproduce the refined mirror principle over the physically relevant range of core masses: the inner and outer boundaries of the envelope evolve in opposite directions. This relationship is therefore a natural consequence of the hydrostatic equilibrium regardless of energy transport. As the envelope’s inner boundary moves inward, its local gravitational acceleration increases, requiring a steeper pressure gradient to maintain hydrostatic equilibrium. Because the pressure at this boundary is fixed by the nuclear-burning conditions of the hydrogen-burning shell, the pressure gradient can steepen only through a reduction in pressure and density in the outer envelope, driving envelope expansion.

\paragraph{(2) The red giant/supergiant phase.} The RG/RSG phase corresponds to a distinct state characterized by an internal structure that differs fundamentally from those of more compact stars. Entry into this phase involves a pronounced structural transition marked by flattening of the temperature gradient, substantial outward redistribution of envelope mass, and the development of an extended convective envelope. Completion of this transition establishes a maximum stellar radius and a characteristic effective temperature of $\sim 4{,}000\,{\rm K}$, defining the formation of RG/RSG. Beyond this point, further inward motion of the envelope’s inner boundary does not produce additional expansion. This saturation occurs because the sharp drop in H$^{-}$ opacity at low temperatures flattens the temperature gradient, modifies the envelope structure, and imposes a physical upper limit on the stellar radius, consistent with the Hayashi limit.

\paragraph{(3) The Hertzsprung gap and red–blue bifurcation.}
Our steady-state solutions further show that, in the intermediate temperature range of $\sim 4,000$-$10,000\,{\rm K}$, a small inward shift of the envelope’s inner boundary can induce a large increase in stellar radius. As a result, stars traverse this yellow regime rapidly and are unlikely to remain there. Instead, they tend to settle into red or blue giant/supergiant configurations. This behavior explains the bifurcation between red and blue supergiants found in grids of stellar models \citep{Ou2023} and accounts for the observed Hertzsprung gap in the HR diagram.

Our steady-state envelope models uncover the key physics underlying RG/RSG formation and complement the MESA stellar evolution results in Paper~I, together providing a unified physical explanation for why stars evolve into red giants or supergiants.


\begin{acknowledgements}
    This research is supported by the National Science and Technology Council, Taiwan, under grant No. NSTC 113-2112-M-001-028-, 114-2811-M-001-094-, 114-2112-M-001-012-, and the Academia Sinica, Taiwan, under a career development award under grant No. AS-CDA-111-M04. This research was supported in part by grant NSF PHY-2309135 to the Kavli Institute for Theoretical Physics (KITP) and grant NSF PHY-2210452 to the Aspen Center for Physics.  KC acknowledges the support of the Alexander von Humboldt Foundation. 
\end{acknowledgements}


\bibliography{RSG}{}
\bibliographystyle{aa}

\end{document}